# Federated Diabetes Prediction in Canadian Adults Using Real-world Cross-Province Primary Care Data

Guojun Tang, MSc[1], Jason E. Black, MSc[2], Tyler S. Williamson, PhD[2,3,4],
Steve H. Drew, PhD, PEng[1]
[1]Department of Electrical and Software Engineering, Schulich School of Engineering, University of Calgary, Calgary, AB; [2]Department of Community Health Sciences, Cumming School of Medicine, University of Calgary, Calgary, AB; [3]O'Brien Institute for Public Health, Calgary, AB; [4]Alberta Children's Hospital Research Institute, Calgary, AB

**Abstract**
*Integrating Electronic Health Records (EHR) and the application of machine learning present opportunities for enhancing the accuracy and accessibility of data-driven diabetes prediction. In particular, developing data-driven machine learning models can provide early identification of patients with high risk for diabetes, potentially leading to more effective therapeutic strategies and reduced healthcare costs. However, regulation restrictions create barriers to developing centralized predictive models. This paper addresses the challenges by introducing a federated learning approach, which amalgamates predictive models without centralized data storage and processing, thus avoiding privacy issues. This marks the first application of federated learning to predict diabetes using real clinical datasets in Canada extracted from the Canadian Primary Care Sentinel Surveillance Network (CPCSSN) without cross-province patient data sharing. We address class-imbalance issues through downsampling techniques and compare federated learning performance against province-based and centralized models. Experimental results show that the federated MLP model presents a similar or higher performance compared to the model trained with the centralized approach. However, the federated logistic regression model showed inferior performance compared to its centralized peer.*

**Introduction**

Predicting diabetes based on patient risk factors is paramount for the Canadian and global populations due to its significant impact on public health and healthcare costs. The number of patients with chronic disease, including diabetes, in Ontario, Canada alone, increased by 11.0% over the 10-year study period to 9.8 million in 2017/18, and the number with multimorbidity increased by 12.2% to 6.5 million[1]. According to Diabetes Canada[2], among Canadians, 30% live with diabetes or prediabetes; 10% live with diagnosed diabetes, a figure that climbs to 15% when cases of undiagnosed type 2 diabetes are included. This high prevalence underscores the importance of prediction and early detection, as managing diabetes early can help prevent complications.

Prediabetes often goes unnoticed and may cause diabetes-related complications, e.g., diabetic kidney disease and retinopathy. Creating an advanced predictive model capable of assimilating patient-specific factors to pinpoint individuals at an elevated risk for diabetes could pave the way for improved preventative strategies at the early stage, significantly reducing expenses associated with hospital stays, drug therapies, and subsequent medical treatments.

Electronic Health Records (EHR) have become a valuable source of patient medical history, opening new avenues for developing innovative data-driven instruments and methodologies to enhance the precision and accessibility of healthcare services. The potential application of machine learning to forecast health outcomes might entail the integration of extensive datasets from disparate health systems. However, developing centralized predictive models requires accessing EHRs from multiple jurisdictions, raising data privacy, confidentiality, and compliance issues. For instance, the Personal Information Protection and Electronic Documents Act (PIPEDA)[3] is a Canadian federal law that applies to the collection, use, and disclosure of personal information in the course of commercial activities in all Canadian provinces as supplemented by substantially similar provincial privacy laws in Alberta, British Columbia, and Québec. Such policies have imposed very restrictive data sharing laws, which form a barrier to data collection in centralized model training. The federated learning may foster collaboration across the medical affiliations from different regions and has been successfully applied in practical medical applications, such as clinical outcome prediction[4] and diagnosis[5].

In response to ongoing data-sharing concerns among healthcare providers, we introduce a federated learning approach that enables the amalgamation of predictive models without storing patient data centrally. A decentralized analytics framework could utilize information from separate health entities without encountering ethical or legal issues. For instance, Humayera et al.[6] used the SUPREME-DM dataset to create a diabetes cohort and employed a federated learning algorithm to develop a prediction model for three distinct diabetic complications. As far as we know, our research represents the inaugural application of federated learning to predict diabetes with authentic clinical datasets. Additionally, predicting diabetes through federated learning presents distinct hurdles due to the naturally skewed distribution in real-life clinical datasets. This research aims to develop a decentralized, privacy-conscious federated learning framework that can breakthrough the mentioned limitations from traditional centralized approaches. We highlight the effectiveness of federated learning in overcoming obstacles associated with employing real-life clinical data. Specifically, we simulate the real-world circumstance by splitting the dataset according to provinces. And we will verify the efficacy of federated learning by comparing the performance of diabetic prediction models trained in the centralized method and the federated method.

The key contributions of our study include the following: (i) we identify the diabetes population using Canadian Primary Care Sentinel Surveillance Network (CPCSSN) [7] as the structured data source; (ii) we develop a decentralized federated learning framework using de-identified EHRs from nine Canadian provinces to predict the diagnosis of diabetes based on the risk factors without sharing patient data cross-province; (iii) we use downsampling techniques to address the class-imbalance issues; (iv) we compare the performance of federated province-based models with centralized models using national-wide data, with and without downsampling.

**Method**
*Data Source*
This study used the Canadian Primary Care Sentinel Surveillance Network (CPCSSN) as the data source. The CPCSSN is a de-identified EHR dataset based on the medical records of nine provinces and over 1.7 million patients. This dataset consists of medical information from practical primary care in Canada, including the patients' demographics, encounters, physical examinations, laboratory test results, medical prescriptions for patients, and disease cases of patients. We defined the cohort with the records in CPCSSN and extracted the features related to diabetes.

*Cohort and Features Selection*
We constructed the cohort according to the instructions from Esser et al.[8]. We chose adult patients who had a blood pressure record in the database between 2004 and 2014 and were aged over 18 at the time of the exam. After that, we removed the patients who had been defined as having diabetes before this period or had medical prescriptions for insulin. Using the blood pressure record as the cohort baseline, we joined the patients' other measurements within the time of the exam no more than one year. Finally, we randomly selected a subset of 30,000 patients from the mentioned records and then applied data cleaning (e.g., filtering out the patient's data with excessive missing features and removing the record with incorrect medical values). The final size of the cohort is 11,631.

**Table 1.** Summary characteristics of the patient information in the cohort. The second column indicates the mean values with standard deviation and range for the continuous variables, or positive cases and their corresponding proportions for the binary variables. The third column displays the number of missing values.

| Characteristics (N = 11,631) | Mean,STD/Frequency (Range/Percentage) | Number of Missing Values |
|---|---|---|
| Age at exam (years) | 57.1, 14.39(18 - 99) | 0 |
| Sex/gender (males) | 5329 (45.82%) | 0 |
| sBP (mmHg) | 126.41, 16.01 (69 -218) | 0 |
| BMI (kg/m$^2$) | 28.94, 6.44 (11 – 66.48) | 2,419 (20.79%) |
| LDL (mmol/L) | 2.82, 0.95 (0 – 8.81) | 51 (0.44%) |
| HDL (mmol/L) | 1.43, 0.45 (0.18 – 6.78) | 3 (0.03%) |
| HbA1c (mg/dL) | 5.936, 0.90 (4 – 17.9) | 3,296 (28.34%) |
| TG (mmol/L) | 1.47, 0.98 (0.2 – 25.57) | 6 (0.05%) |
| Hypertension | 4,595 (39.51%) | 0 |
| Depression | 2,489 (21.40%) | 0 |
| Osteoarthritis | 2,101 (18.06%) | 0 |
| COPD | 617 (5.27%) | 0 |
| Hypertension Med | 4,989 (42.89%) | 0 |
| Corticosteroids | 151 (12.98%) | 0 |
| Diabetes | 2,135 (18.36%) | 0 |

We chose the following features as the predictive variables[6]: age at the exam, gender, systolic blood pressure (sBP), body mass index (BMI), low-density lipoprotein (LDL), high-density lipoprotein (HDL), HemoglobinA1c (HbA1c), triglycerides (TG), depression, hypertension, medications for hypertension, osteoarthritis, chronic obstructive pulmonary disease (COPD), and corticosteroids. Table 1 shows the summary of the predictive variables. For the definition of hypertension medications, we adopted the criteria from Garies et al.[9]. In this study, we considered the definition of the disease cases (i.e., diabetes) using the *DiseaseCase* table in the CPCSSN database, in which these case definitions are verified by calculating the sensitivity and specificity of these definitions compared to chart review by an expert EMR user[10].

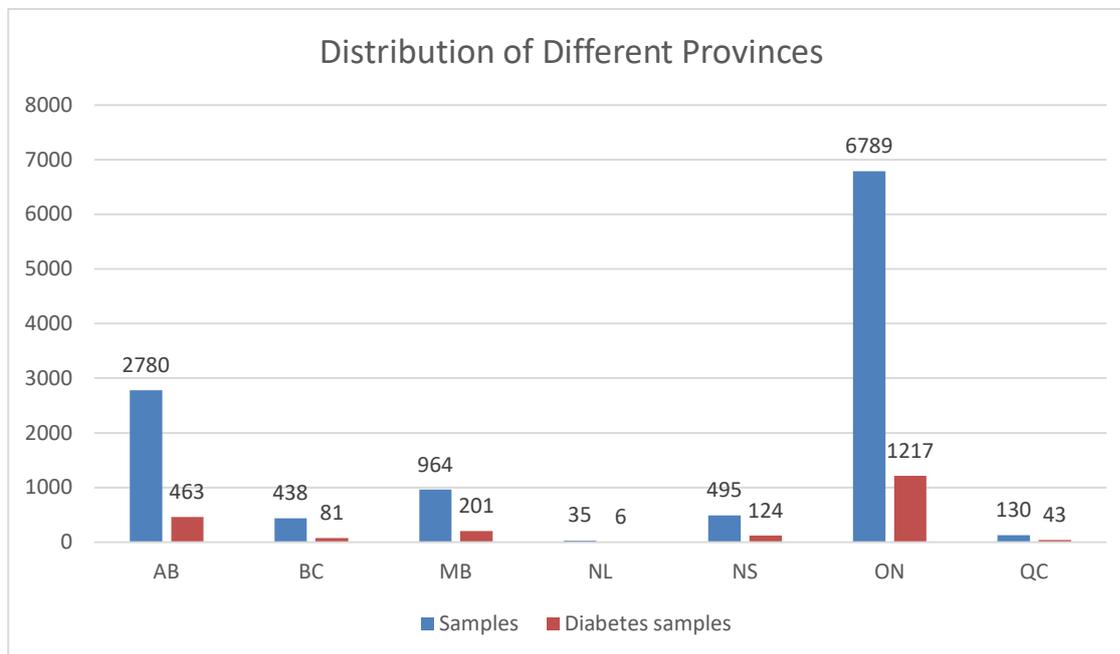

**Figure 1.** The statistics of the total samples and the positive samples among different provinces.

*Centralized Machine Learning*
To illustrate the efficacy of how federated learning improves the performance of the predictive model, first, we partitioned the dataset into different subsets according to the patients' provinces. We trained the provincial model, or the local model, by its corresponding provincial data under the assumption that a province can access the patient data within its boundaries without regulatory barriers. Therefore, we may simulate the isolation of real-world data among those regions. Then, we shuffle the local dataset and preserve 70% and 30% of the data for model training and testing, respectively. There are missing values in the variables, as shown in Table 1, and we imputed those missing values by applying the MICE algorithm[11] to the training and testing set separately. After we trained all local models, we aggregated each training set and test set for training a centralized model, which can be regarded as the ideal baseline of the federated model. Those local and centralized models act as the benchmark and are compared to the federated model. We choose centralized machine learning as a benchmark because it always acquires more comprehensive information and usually has better performance compared to FL. Using local models as the other benchmark may help us verify whether FL improves the performance of the model. We employed Scikit-learn[12] to implement the logistic regression and multiple layer perceptron (MLP) algorithms, and the models' hyperparameters are shown in Table 2. The statistics of each local dataset are shown in Figure 1.

**Table 2.** Model hyperparameters of **Logistic Regression** and **MLP** in centralized machine learning and federated learning.

| Model | Hyperparameters | Values |
| --- | --- | --- |
| Logistic Regression | Penalty<br>Solver<br>C | L1<br>liblinear<br>1.0 |
| MLP | Learning rate<br>Alpha<br>Activation<br>Solver<br>Hidden Layers sizes | 0.001<br>0.01<br>relu<br>adam<br>(128, 128) |

*Federated Learning*
Federated learning (FL) is a distributed machine learning paradigm that empowers different clients who have sensitive private data to collaboratively train a global model by uploading their local model parameters instead of sharing the data directly. The overview of FL is illustrated in Figure 2. In our experiment, we adopted the FedAvg[13] algorithm to implement federated logistic regression and federated MLP by using Scikit-learn[12] and Flower[14], a lightweight federated learning framework in Python. The FedAvg algorithm first dispatches the initialized model to all clients and launches multiple global training rounds to train the global model. In each global training round, the central server randomly selects $n$ clients involved within this round of training. The selected clients updated the local model by computing the gradient using the global parameters and their own data after downloading the global model parameters from the central server. In the local update step, the client updated their local model in $E$ local epochs and $b$ local batch size with their local data. After the local update step, the central server receives clients' model parameters. It aggregates them by weighted averaging the parameters (we consider equal weights in this case), which will be broadcast to all clients at the end of this global training round. The overview of the FedAvg algorithm is shown in Algorithm 1.

**Algorithm 1.** The pseudocode overview of FedAvg[13].

---

**Algorithm 1 FedAvg**. *K* total clients; *n* participants each training round; *T* global epoch; *E* local epoch; *l* loss function; *b* batch size; $\eta$ learning rate;

---

    **procedure** ServerAggregation initialize model parameter $w_0$
        **for** $t = 1,2...,T$ **do**
            $S_t \leftarrow$ randomly choose $n$ clients **for** $k \in S_t$ **parallel do**
                $w_{t+1}^k \leftarrow$ ClientUpdate($k, w_t$)
        $w_{t+1} \leftarrow \frac{|S_t|}{K} \sum_{k \in S_t} w_{t+1}^k$
    **procedure** ClientUpdate($k, w$)

```
for i = 1,...,E do
    for local batch b ∈ dataset from client k do
        w ← w − η∆l(w;b)
return w
```

For the FL setting, we employed the same model hyperparameters as centralized machine learning. We divided the dataset into seven subsets according to the provinces, including Alberta (AB), British Columbia (BC), Manitoba (MB), Newfoundland and Labrador (NL), Nova Scotia (NS), Ontario (ON), and Quebec (QC). New Brunswick (NB) and Prince Edward Island (PEI) are not involved due to lack of samples. These constructed seven clients are equipped with the corresponding provincial local data and trained a global model using the FedAvg algorithm. In this experiment, we set the participants $n = 2$ while the local epoch $E = 1$.

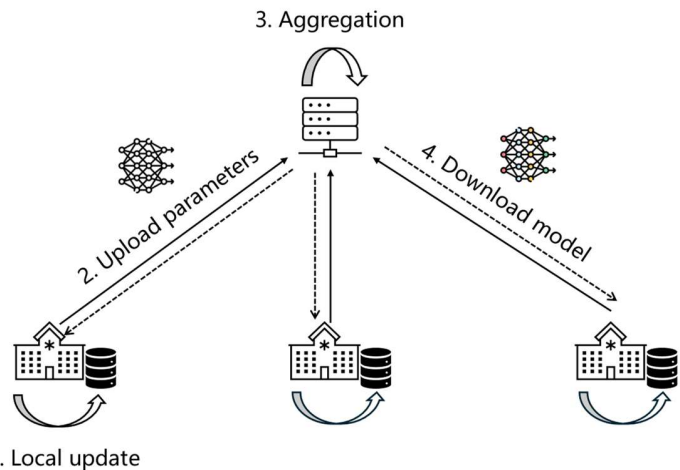

**Figure 2.** Overview of federated learning framework.

## Results
### *Experimental Settings*
In our experiment, we trained the federated model using the FedAvg algorithm with data from different provinces and compared this model with the provincial local and centralized models. As the disease prediction model is a binary classification task, we considered the F1 scores and AUC scores as benchmarks of the models, which may reflect the efficiency of the binary classification rather than the accuracy. Furthermore, we also plotted the calibration curves for these models in order to evaluate the class probabilities from them. Due to the nature of class imbalance shown in Figure 1, we adopted dataset downsampling, a common strategy for imbalanced datasets, and tested its efficiency in FL. We conducted our experiment in the CPCSSN Secure Research Environment (SRE), equipped with Microsoft Windows 10 OS, Intel Xeon Silver 4216 CPU, and 32GB of RAM.

### *Experimental Results*
We conducted the experiment in which we first trained the performance of logistic regression and MLP using local data among different provinces, a centralized dataset, and federated machine learning methods. The summaries of the results are presented in Table 3 and Table 4, respectively. The table recorded the global test set results of AUC, F1, precision, and recall by applying different types of models: 1) local models trained by the provincial data from its own region (AB, BC, MB, NL, NS, ON, QC); 2) the centralized model trained by the data from all regions (CML); 3) the federated model trained using the FedAvg (FL). The table recorded the global test set results of AUC, F1, precision, and recall by applying different types of models: 1) local models trained by the provincial data from its own region (AB, BC, MB, NL, NS, ON, QC); 2) the centralized model trained by the data from all regions (CML); 3) the federated model trained using FedAvg (FL). We also considered the training strategies with downsampling and without resampling. The performance (F1-score and recall) of the local models is roughly consistent with the data quality (e.g., the number and the proportion of the positive samples) in both logistic regression and MLP models. The FL method may empower us to improve the overall model performance of those areas with relatively low quality by averaging the parameters from the models trained by the regional dataset with better quality.

It is worth noting that, however, the performance of logistic regression and MLP varies under the FL framework. The performance (AUC, F1, precision, and recall) of federated MLP only differed from approximately 3% and even surpassed the results from the centralized model. However, in the federated logistic regression model, the AUC was reduced by 13% in the no-resampling case and 8% in the downsampling case when compared to the centralized method. This condition also has been incurred in the F1 score, reduced by 20% in the no-resampling case and 16% in the downsampling case.

Notably, the impact of the downsampling will be more significant in the MLP model. After applying the downsampling, the AUC increased by 3% and 8% in the centralized and federated logistic regression models, respectively, while the F1 decreased by 6% in CML and 2% in FL. However, the downsampling will cause a 10% decrease in the F1 in centralized MLP and an 8% decrease in federated MLP, in which we could only observe a 2% improvement in the AUC.

**Table 3.** Comparison of the **Logistic Regression** performance of the local models, centralized model (CML), and federated model (FL) using the global test dataset.

| Model | | AUC | F1 | Precision | Recall |
|---|---|---|---|---|---|
| AB | No resample | 0.8702 | 0.8194 | 0.8856 | 0.7625 |
|    | Downsample  | 0.8827 | 0.7275 | 0.6148 | 0.8906 |
| BC | No resample | 0.8315 | 0.7730 | 0.8934 | 0.6812 |
|    | Downsample  | 0.8806 | 0.7296 | 0.6225 | 0.8812 |
| MB | No resample | 0.8579 | 0.8125 | 0.9140 | 0.7312 |
|    | Downsample  | 0.8754 | 0.7462 | 0.6679 | 0.8453 |
| NL | No resample | 0.8241 | 0.7579 | 0.8719 | 0.6703 |
|    | Downsample  | 0.8702 | 0.6937 | 0.5668 | 0.8937 |
| NS | No resample | 0.8696 | 0.8092 | 0.8541 | 0.7687 |
|    | Downsample  | 0.8368 | 0.6264 | 0.4850 | 0.8843 |
| ON | No resample | 0.8703 | 0.8262 | 0.9082 | 0.7578 |
|    | Downsample  | 0.8878 | 0.7729 | 0.7043 | 0.8562 |
| QC | No resample | 0.8728 | 0.7276 | 0.6317 | 0.8578 |
|    | Downsample  | 0.8288 | 0.6013 | 0.4499 | 0.9062 |
| CML | No resample | 0.8686 | 0.8217 | 0.8996 | 0.7562 |
|     | Downsample  | 0.8909 | 0.7657 | 0.6817 | 0.8734 |
| FL | No resample | 0.7388 | 0.6211 | 0.8075 | 0.5046 |
|    | Downsample  | 0.8134 | 0.6041 | 0.4730 | 0.8359 |

**Table 4.** The comparison table of the **MLP** performance of the local models, centralized model (CML), and the federated model (FL), using the global test dataset.

| Model | | AUC | F1 | Precision | Recall |
|---|---|---|---|---|---|
| AB | No resample | 0.8379 | 0.7195 | 0.6901 | 0.7515 |
|  | Downsample | 0.8411 | 0.6527 | 0.5285 | 0.8531 |
| BC | No resample | 0.8167 | 0.6942 | 0.6812 | 0.7078 |
|  | Downsample | 0.8228 | 0.6173 | 0.4856 | 0.8468 |
| MB | No resample | 0.8348 | 0.7453 | 0.7775 | 0.7156 |
|  | Downsample | 0.8371 | 0.6540 | 0.5371 | 0.8359 |
| NL | No resample | 0.7250 | 0.5786 | 0.6807 | 0.5031 |
|  | Downsample | 0.7817 | 0.5376 | 0.3889 | 0.8703 |
| NS | No resample | 0.8431 | 0.7298 | 0.7039 | 0.7578 |
|  | Downsample | 0.8102 | 0.5897 | 0.4502 | 0.8546 |
| ON | No resample | 0.8568 | 0.7827 | 0.8166 | 0.7515 |
|  | Downsample | 0.8559 | 0.6789 | 0.5591 | 0.8640 |
| QC | No resample | 0.8175 | 0.6376 | 0.5349 | 0.7890 |
|  | Downsample | 0.8038 | 0.5777 | 0.4359 | 0.8562 |
| CML | No resample | 0.8499 | 0.7759 | 0.8205 | 0.7359 |
|  | Downsample | 0.8496 | 0.6741 | 0.5585 | 0.8500 |
| FL | No resample | 0.8665 | 0.8176 | 0.8942 | 0.7531 |
|  | Downsample | 0.8808 | 0.7380 | 0.6399 | 0.8718 |

When clients are sharing models, the question comes up: should the models be aggregated or is it effective enough to apply the model with relatively good performance? To validate the generalization of FL and the necessity of the model aggregation, we tested the federated model with the local test sets from the less-sampled regions (BC, MB, NS, and QC) and compared the performance with the centralized model and the local models, which had relatively adequate training data (AB and ON). The results are shown in Table 5 and Table 6. While using logistic regression, the performance of the AB and ON local models surpassed the models using centralized machine learning and federated learning in most cases. However, the AUC and F1 of the federated MLP model perform closely or even better than the model of those two regions with a larger amount of data samples.

**Table 5.** The comparison table of the **Logistic Regression** performance of the local models with adequate data (AB, and ON), centralized model (CML), and the federated model (FL), using the test data from less-sampled regions (BC, MB, NS, and QC).

| Model | | BC-TEST | | MB-TEST | | NS-TEST | | QC-TEST | |
|---|---|---|---|---|---|---|---|---|---|
|  |  | AUC | F1 | AUC | F1 | AUC | F1 | AUC | F1 |
| AB | No resample | 0.7569 | 0.6500 | 0.8786 | 0.8318 | 0.9099 | 0.8857 | 0.7500 | 0.6667 |
|   | Downsample | 0.8078 | 0.6667 | 0.8913 | 0.7659 | 0.9060 | 0.8045 | 0.7115 | 0.6086 |
| ON | No resample | 0.7569 | 0.6500 | 0.8786 | 0.8318 | 0.8918 | 0.8787 | 0.7692 | 0.7000 |
|   | Downsample | 0.7800 | 0.6521 | 0.8920 | 0.7938 | 0.9103 | 0.8292 | 0.7307 | 0.6364 |
| CML | No resample | 0.7569 | 0.6500 | 0.8786 | 0.8318 | 0.8874 | 0.8656 | 0.7692 | 0.7000 |
|   | Downsample | 0.7800 | 0.6521 | 0.8960 | 0.7910 | 0.9104 | 0.8139 | 0.7307 | 0.6364 |
| FL | No resample | 0.6064 | 0.3529 | 0.7786 | 0.6930 | 0.7433 | 0.6440 | 0.7307 | 0.6315 |
|   | Downsample | 0.8101 | 0.6428 | 0.8184 | 0.6375 | 0.8432 | 0.7096 | 0.7115 | 0.6153 |

**Table 6.** The comparison table of the **MLP** performance of the local models with adequate data (AB, and ON), centralized model (CML), and the federated model (FL), using the test data from less-sampled regions (BC, MB, NS, and QC).

| Model |   | BC-TEST |   | MB-TEST |   | NS-TEST |   | QC-TEST |   |
|---|---|---|---|---|---|---|---|---|---|
|   |   | AUC | F1 | AUC | F1 | AUC | F1 | AUC | F1 |
| AB | No resample | 0.7569 | 0.6500 | 0.8670 | 0.7656 | 0.8608 | 0.7654 | 0.6730 | 0.5454 |
|   | Downsample | 0.8032 | 0.6538 | 0.8485 | 0.6887 | 0.8297 | 0.6956 | 0.6153 | 0.5000 |
| ON | No resample | 0.7685 | 0.6511 | 0.8992 | 0.8474 | 0.9011 | 0.8421 | 0.8076 | 0.7500 |
|   | Downsample | 0.8032 | 0.6538 | 0.8764 | 0.7412 | 0.8521 | 0.7252 | 0.7500 | 0.6667 |
| CML | No resample | 0.7361 | 0.6153 | 0.8677 | 0.7966 | 0.9099 | 0.8857 | 0.7115 | 0.6000 |
|   | Downsample | 0.7476 | 0.5660 | 0.8800 | 0.7284 | 0.8835 | 0.7727 | 0.6153 | 0.5161 |
| FL | No resample | 0.7569 | 0.6500 | 0.8807 | 0.8392 | 0.9144 | 0.8985 | 0.7115 | 0.6000 |
|   | Downsample | 0.7916 | 0.6530 | 0.9126 | 0.8088 | 0.8700 | 0.7586 | 0.7115 | 0.6086 |

Figure 3 showed the calibration curves of the CML and FL models, tested by the global test set. Due to the privacy concern that we cannot share the data, we did not consider the model recalibration. MLP models tended to underestimate the risk compared to logistic regression models in both CML and FL settings. After applying the FL approach, models will generally overestimate the risk, especially if the mean predicted probability exceeds 0.5. The CML and FL logistic regressions showed good calibration, with curves that were relatively close to the diagonal. The FL MLP, which tended to overestimate the risk compared to the CML model, had a calibration curve closer to the diagonal and indicated a better calibration. Once we adopt the downsampling, the curve may significantly underestimate the risk, leading to poor calibration.

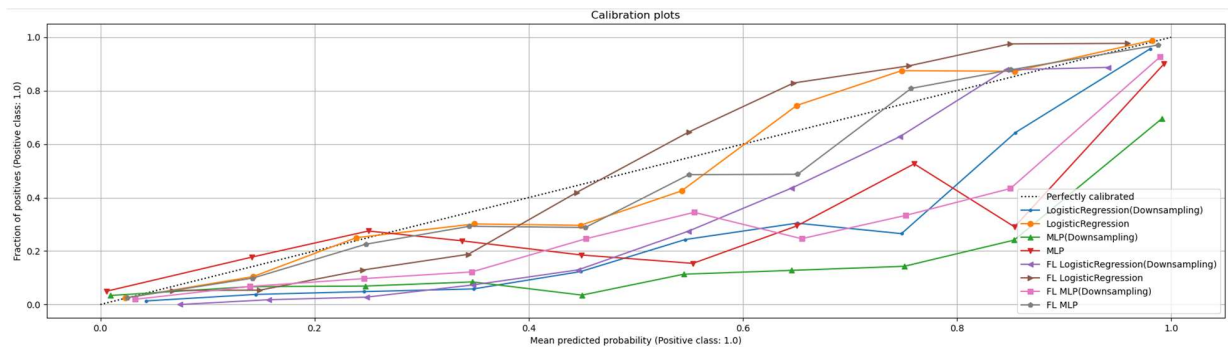

**Figure 3.** Calibration curves of CML models and FL models.

## Discussion

Table 3 illustrated that the federated logistic regression model underperformed its centralized model. The experimental results from Humayera[4] also indicated a similar drop in performance between the centralized logistic model and the federated model. A reasonable explanation is that the model parameters of the neural network, such as MLP, are more complicated than the linear model, such as logistic regression. Therefore, it may still preserve more features after averaging the weights of all the local clients.

According to the conclusion from Ruben et.al.[15], the training with downsampling may have a higher recall than the datasets without resampling. However, it will also significantly downgrade precision, leading to a worse F1 score. This trend has been seen across all local, centralized, and federated models in our studies. Moreover, by comparing the outcomes of Table 3 and Table 4, it becomes noticeable that the reduction in performance due to the downsampling strategy will be more considerable in federated MLP.

Based on the results in Table 5 and Table 6, it can be inferred that the federated MLP model may be more robust and outperform the local models trained by regions with high-quality data. However, the federated logistic regression underperformed those local models. The potential interpretation is that the federated learning framework, which aggregates various local models, may extract more features from the data compared to the local model only trained by the data from a single region. In practice, while using the logistic regression model, directly applying the local model from the region with good data quality is an alternative solution instead of training a federated model.

The calibration curves of the centralized and federated models are shown in Figure 3. Both the centralized and federated logistic regression models produced well-calibrated predictions due to the canonical link function in the linear model[16]. Following the implementation of the Federated Learning (FL) approach, the MLP model demonstrated a tendency to generally overestimate predictions and showed better calibration in comparison to its centralized version. We noticed that the implementation of downsampling leads to decreased calibration performance in both federated and centralized models.

## Conclusion and Future Work

Federated learning is an innovative solution to train a global model in a distributed system without sharing the data directly or violating the privacy policy. It allows us to break the barrier of privacy restrictions among different regions and improve the model quality. In this paper, we applied FedAvg to a diabetes prediction task and validated its effectiveness by comparing it with the local and centralized models. Our result indicates that the federated MLP model has a favorable impact on the model quality rather than only using the local or centralized models. However, as for the linear model of logistic regression, even though it has been widely used in assorted medical scenarios, the performance, which has been shown in our experiments, is still considerably lower than the centralized model, which has been commonly regarded as the baseline of FL.

FL has provided a feasible distributed machine learning framework for privacy-sensitive data and has a great potential in the healthcare scenario. From an intuitive perspective, the performance of the centralized machine learning model may be regarded as the upper bound of FL. However, the empirical results from our study and the relevant research from Humayera et al.[4] indicated that the federated MLP outperformed the centralized one. In our future works, we may corroborate the results and attempt to figure out an explanation for this phenomenon. Our studies only focus on two specific models: logistic regression and MLP. Our future research will involve the exploration of more model types, including XGBoost[17], a tree-based machine learning model that has been widely applied in various medical applications. The model calibration technique is a very common machine learning technique in healthcare applications, but we did not consider it in our study. Our future work will also include the model calibration in FL. In this study, we only adopted the simple case of predicting the diagnosis of diabetes. In the future, based on the Canadian patients and CPCSSN dataset, we may introduce FL into further research of diabetes-related complications, such as chronic kidney diseases.

## Acknowledgements
This material is based in part upon work supported by Natural Sciences and Engineering Research Council of Canada (NSERC) Discovery Grant Program (RGPIN-2024-03954).

# References

1. Steffler M, Li Y, Weir S, Shaikh S, Murtada F, Wright JG, Kantarevic J. Trends in prevalence of chronic disease and multimorbidity in Ontario, Canada. Cmaj. 2021 Feb 22;193(8):E270-7.
2. Diabetes in Canada: Backgrounder. Ottawa: Diabetes Canada. 2023.
3. Legislative Services Branch. Personal Information Protection and Electronic Documents Act. Justice.gc.ca. 2019. Available from: https://laws-lois.justice.gc.ca/eng/acts/P-8.6/
4. Dayan, I., Roth, H. R., Zhong, A., Harouni, A., Gentili, A., Abidin, A. Z., ... & Li, Q. (2021). Federated learning for predicting clinical outcomes in patients with COVID-19. Nature medicine, 27(10), 1735-1743.
5. Qayyum, A., Ahmad, K., Ahsan, M. A., Al-Fuqaha, A., & Qadir, J. (2022). Collaborative federated learning for healthcare: Multi-modal covid-19 diagnosis at the edge. IEEE Open Journal of the Computer Society, 3, 172-184.
6. Islam H, Mosa A, Famia. A Federated Mining Approach on Predicting Diabetes-Related Complications: Demonstration Using Real-World Clinical Data. AMIA Annu Symp Proc. 2022;2021:556-564. Published 2022 Feb 21.
7. Birtwhistle R, Keshavjee K, Lambert-Lanning A, et al. Building a pan-Canadian primary care sentinel surveillance network: initial development and moving forward. The Journal of the American Board of Family Medicine, 2009, 22(4): 412-422.
8. Esser K, Duong M, Kain K, Tran S, Sadeghi A, Aziz Guergachi, et al. Predicting Diabetes in Canadian Adults Using Machine Learning. medRxiv (Cold Spring Harbor Laboratory). 2024 Feb 5;
9. Garies S, Hao S, McBrien K, Williamson T, Peng M, Khan NA, et al. Prevalence of Hypertension, Treatment, and Blood Pressure Targets in Canada Associated With the 2017 American College of Cardiology and American Heart Association Blood Pressure Guidelines. JAMA Network Open. 2019 Mar 8;2(3):e190406.
10. Williamson T, Green ME, Birtwhistle R, Khan S, Garies S, Wong ST, et al. Validating the 8 CPCSSN Case Definitions for Chronic Disease Surveillance in a Primary Care Database of Electronic Health Records. The Annals of Family Medicine. 2014 Jul 1;12(4):367–72.
11. Buuren S van, Groothuis-Oudshoorn K. mice: Multivariate Imputation by Chained Equations inR. Journal of Statistical Software. 2011;45(3).
12. Scikit-learn: Machine Learning in Python, Pedregosa et al., JMLR 12, pp. 2825-2830, 2011.
13. McMahan B, Moore E, Ramage D, et al. Communication-efficient learning of deep networks from decentralized data. Artificial intelligence and statistics. PMLR, 2017: 1273-1282
14. Beutel D J, Topal T, Mathur A, et al. Flower: A friendly federated learning research framework. arXiv preprint arXiv:2007.14390, 2020.
15. van den Goorbergh R, van Smeden M, Timmerman D, Van Calster B. The harm of class imbalance corrections for risk prediction models: illustration and simulation using logistic regression. Journal of the American Medical Informatics Association. 2022 Jun 10;
16. Mario V. Wüthrich, Michael Merz (2023). "Statistical Foundations of Actuarial Learning and its Applications" Springer Actuarial
17. Chen T, Guestrin C. XGBoost: a Scalable Tree Boosting System. Proceedings of the 22nd ACM SIGKDD International Conference on Knowledge Discovery and Data Mining - KDD '16. 2016;785–94.